\documentclass[aps,prl,twocolumn,superscriptaddress,showpacs]{revtex4}
\usepackage{amssymb}
\usepackage{epsfig}
\usepackage{amsmath}
\usepackage{times}

\begin{document}
\title{Explosive synchronization in adaptive and multilayer networks}

\author{Xiyun Zhang}
\affiliation{Department of Physics, East China Normal University,
Shanghai, 200062, China}

\author{Stefano Boccaletti}
\email{stefano.boccaletti@gmail.com}
\affiliation{CNR- Institute of Complex Systems, Via
Madonna del Piano 10, 50019 Sesto Fiorentino, Florence, Italy}
\affiliation{The Italian Embassy in Israel, 25 Hamered Street, 68125 Tel Aviv, Israel}

\author{Shuguang Guan}
\affiliation{Department of Physics, East China Normal University,
Shanghai, 200062, China}

\author{Zonghua Liu}
\email{zhliu@phy.ecnu.edu.cn} \affiliation{Department of Physics, East China Normal University,
Shanghai, 200062, China}

\date{\today}

\begin{abstract}

Explosive synchronization (ES) is nowadays a hot topic of interest in nonlinear science and
complex networks. So far, it is conjectured that ES is rooted in the setting of
specific microscopic correlation features between the natural frequencies of the networked oscillators
and their effective coupling strengths.
We show that ES, in fact, is far more general, and can occur in adaptive and multilayer networks
also in the absence of such correlation properties. Precisely, we first report evidence of ES in the absence of correlation
for networks where a fraction $f$ of the nodes have links adaptively controlled by a local order parameter, and
then we extend the study to a variety of two-layer networks with a
fraction $f$ of their nodes coupled each other by means of dependency links. In this latter case, we even show that ES sets in,
regardless of the differences in the frequency distribution and/or in the topology of connections between the two layers.
Finally, we provide a rigorous, analytical, treatment to properly ground all the observed scenario, and to facilitate the understanding of
the actual mechanisms at the basis of ES in real-world systems.

\end{abstract}


\maketitle

Phase transitions in ensembles of networked systems is one of the hottest topics of interest in current days.
Recently, it was pointed out that the transition of an ensemble of networked phase oscillators from incoherence to synchronization
can be first-order like, discontinuous and irreversible, called explosive synchronization (ES).
This discovery is of huge significance, as abrupt phase transitions are indeed
featured by a variety of real world systems \cite{bocca}: from epileptic seizures in the brain
\cite{Dhamala:2013}, to cascading failure of power grids \cite{Buldyrev:2010} and jamming in the Internet
\cite{Huberman:1997}. Since its finding in 2005 \cite{pazo:2005}, ES has been paid a great attention
\cite{Jesus:2011,Leyva:2012,Peron:2012,Leyva:2013a,Zhang:2013,Leyva:2013b,Li:2013,Zhang:2014}.
For instance, it was studied in the context of periodic phase oscillators for scale-free (SF) network's topologies, with an ad-hoc imposed
positive correlation between the natural frequencies of the oscillators and the degrees of nodes
\cite{Jesus:2011}, and the experimental verification of such a setup was given in an electronic circuits with a star
configuration and chaotic units \cite{Leyva:2012}. Later on, ES was described for generic network's topologies (either SF or non-SF)
in a modified Kuramoto model, with a positive correlation
between the natural frequencies of oscillators and their coupling strengths \cite{Zhang:2013,Leyva:2013b}.
The two kinds of positive correlations can be unified into the framework of effective couplings in
mean-field, which are weighted to be proportional to the frequency (or the frequency difference) of the oscillators
\cite{Leyva:2013a,Zhang:2013}. More recently, it was also shown that ES can be considered as the counterpart of an explosive
percolation process in dynamical phase space \cite{Zhang:2014}.

The accepted state of knowledge on this matter is, hence, that ES has a basic and key microscopic root in the setting of local correlation features
(either ad-hoc imposed \cite{Jesus:2011,Leyva:2012}, or
spontaneously emerging \cite{Peron:2012,Leyva:2013a,Zhang:2013,Leyva:2013b,Li:2013,Zhang:2014}) between the natural frequency of a networked
oscillator and its degree, or effective coupling strength.
In this Letter, we fundamentally revisit the issue, and provide an answer to the following question: Is
it possible to observe ES in networked oscillators {\it without} the presence of {\it any kind} of microscopic correlation features?
We first consider a network where the coupling of a fraction $f$ of the nodes is adaptively controlled by a local
order parameter, and show that ES emerges, indeed, when the value of $f$ is over a critical value $f_c$.
We then extend the study of ES to multilayer networks. Precisely, we give evidence that ES
is a generic feature of two-layered networks, when a fraction $f$ of their
nodes are coupled with each other by means of {\it dependency} links, i.e. when the coupling strength
of a node in a layer is adaptively controlled by the local order
parameter of the corresponding node in the other layer. Further, we present a rigorous theoretical analysis of
mean-field to account for all the described scenario. Finally, we formulate the main conclusion:
the robustness of our findings
suggests that all previous studies on ES can be, in fact, unified into a common root, that of suppressing the
formation of giant clusters.

Let us begin with considering a network of $N$ Kuramoto-like phase oscillators, with an explicit fraction
$f$ of the nodes adaptively controlled by a local order parameter. In our model, the evolution of each oscillator is ruled by
\begin{equation}
\dot{\theta}_{i}=\omega_{i}+\lambda \alpha_{i}\sum_{j=1}^{N}A_{ij}\sin(\theta_{j}-\theta_{i})
\label{Kuramoto1}
\end{equation}
where $i=1,\cdots, N$, $\omega_i$ ($\theta_i$) is the natural frequency (the instantaneous phase) of the  $i^{th}$
oscillator, $\lambda$ is the overall coupling strength, $k_i=\sum_{j=1}^N A_{ij}$ is the degree of
node $i$, and $A_{ij}$ are the elements of the the network's adjacency matrix $A$ ($A_{ij}=1$ when the nodes $i$ and
$j$ are connected, and $A_{ij}=0$ otherwise). When compared to previous studies, the key feature of
Eq. (\ref{Kuramoto1}) is the presence of the parameter $\alpha_{i}$. To define $\alpha_i$, we introduce the instantaneous local order
parameter for the $i^{th}$
oscillator of the network, as
\begin{equation}
\label{local-order1}
r_{i}(t)e^{i\phi}=\frac{1}{k_{i}}\sum_{j=1}^{k_{i}} e^{i\theta_{j}}.
\end{equation}
By definition, $0\leq r_{i}\leq 1$, and $\phi$ denotes the average (over the ensemble of neighbors) phase.
Then, we randomly choose a fraction $f$ of network's nodes, and set for all of them $\alpha_{i}=r_{i}$. The
remaining fraction $1-f$ of nodes will have instead, $\alpha_{i}=1$. That is, the fraction $f$ of nodes is actually
adaptively controlled by the corresponding local order parameters.

\begin{figure}
\epsfig{figure=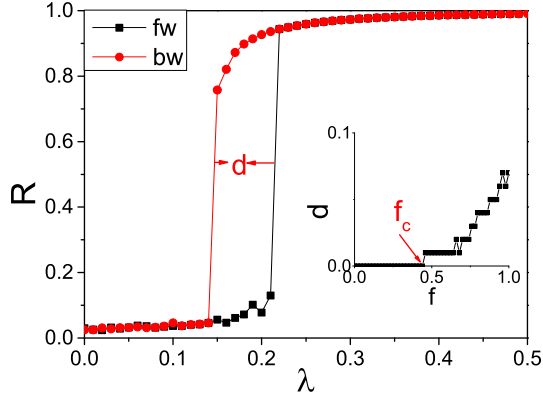,width=1.0\linewidth} \caption{(color
online). Forward (black line with squares) and backward (red line with circles) synchronization
transitions for a single network with $N=1,000$ and $f=1$. The inset reports the dependence of $d$ on $f$.
See text for the specifications on the network topology and on the frequency distribution. }
\label{Fig:single-network}
\end{figure}

The degree of phase coherence in the network can be measured by means of the global order parameter $R$ defined by
\begin{equation}
\label{global-order1}
Re^{i\Psi}=\frac{1}{N}\sum_{j=1}^N e^{i\theta_{j}}
\end{equation}
where $0\leq R \leq 1$ and $\Psi$ denotes the average phase. In these first numerical simulations, we draw the set of
frequencies $\{ \omega_{i} \}$ in Eq. (\ref{Kuramoto1}) from a random homogeneous distribution
$g(\omega)$ in the range $[-1,1]$, and we consider a random Erd\H{o}s-R\'{e}nyi (ER)
network \cite{BA:2002} with size $N=1,000$ and  average degree
$\langle k\rangle=\frac{1}{N}\sum_{i=1}^Nk_{i}=12$. We increase (decrease) the coupling strength $\lambda$
adiabatically with an increment (decrement) $\delta\lambda=0.01$ from $\lambda=0$ ($\lambda=0.5$) and compute the stationary value
of $R$ for each $\lambda$ \cite{details} during the forward (backward) transition from the incoherent to the phase synchronized state.
Fig. \ref{Fig:single-network} reports $R$ vs. $\lambda$ for the case of $f=1$, with the ``squares" and ``circles" labeling  the curves of
the forward and backward transitions, respectively. It is evident the presence of an abrupt transition
with an associated hysteretic loop in $R$, indicating the occurrence of ES in Eq. (\ref{Kuramoto1}).
Denoting by $d$ the width of such an hysteretic loop, the inset of Fig. \ref{Fig:single-network}
reports the dependence of $d$ on $f$, and shows the existence of a critical value $f_c$ where $d$ passes from being zero (i.e. a second-order phase transition)  to a finite value.

Our second step is showing the generality of our findings for two-layered networks, with different topological and
frequency configurations. To this purpose, we construct two independent networks (I and II) and let them
have the same size $N$ in such a way that the nodes on the two layers have a one-to-one correspondence
(in the following we denote, for simplicity, each pairs of nodes by the same index $i$).
We again assume that a fraction $f$ of the nodes between the two networks are coupled with each other by forming dependency
links \cite{Parshani:2011}.
The equations of motion can be written as
\begin{eqnarray}
\dot{\theta}_{i,1}&=&\omega_{i,1}+\lambda \alpha_{i,1}\sum_{j=1}^{k_{i,1}}\sin(\theta_{j,1}-\theta_{i,1}), \nonumber \\
\dot{\theta}_{i,2}&=&\omega_{i,2}+\lambda \alpha_{i,2}\sum_{j=1}^{k_{i,2}}\sin(\theta_{j,2}-\theta_{i,2}),
\label{Kuramoto}
\end{eqnarray}
where $i=1,\cdots, N$ and the subscripts $1,2$ stand for the layers I and II, respectively.
In Eqs. (\ref{Kuramoto}), the average degree is $\langle k_1\rangle=\frac{1}{N}\sum_{i=1}^Nk_{i,1}$ ($\langle k_2\rangle=\frac{1}{N}\sum_{i=1}^Nk_{i,2}$)
for the layer I (II), and the parameters
$\alpha_{i,1}$ and $\alpha_{i,2}$ account for the coupling
between the two layers. Precisely, we set $\alpha_{i,1}=r_{i,2}$ and $\alpha_{i,2}=r_{i,1}$ if the
pair of nodes $i$ is part of the fraction $f$ of coupled nodes (otherwise we set
$\alpha_{i,1}=\alpha_{i,2}=1$), where $r_{i,1}$ and $r_{i,2}$ are defined by $
r_{i,1}e^{i\phi_1}=\frac{1}{k_{i,1}}\sum_{j=1}^{k_{i,1}} e^{i\theta_{j,1}}$ and
$r_{i,2}e^{i\phi_2}=\frac{1}{k_{i,2}}\sum_{j=1}^{k_{i,2}} e^{i\theta_{j,2}}$.
In this way, a group of oscillators in layer I is adaptively controlled
by the local order parameters of the corresponding nodes on layer II, and vice-versa.

\begin{figure}
\epsfig{figure=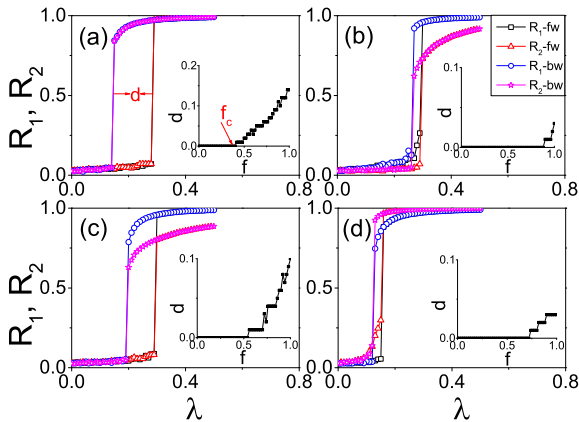,width=1.0\linewidth} \caption{(color
online). Synchronization transitions in two-layer networks for $N=1,000$ and $f=1$. In all plots, ``squares" and
``circles" ("triangles" and "stars") are used for reporting the values of $R_1$ ($R_2$), ad the insets
show the corresponding dependence of $d$ on $f$. Layer I is fixed as a random ER network with average
degree $\langle k_1\rangle=12$, and having a random homogeneous distributions of frequencies in the
range $[-1,1]$. Layer II has different specifications as follows: (a) it is another ER network
with $\langle k_2\rangle=12$, and $g(\omega_{i,2})$ is a random homogeneous distribution in the range $[-1,1]$;
(b) it is an ER network with $\langle k_2\rangle=6$ and $g(\omega_{i,2})$ is the same as
in (a); (c) it is the same as in (a) but $g(\omega_{i,2})$ is now a Lorentzian distribution (see text for definition)
with $\omega_0=0$ and $\gamma=0.5$; (d) it is a BA network with $\langle k_2\rangle=12$
and $g(\omega_{i,2})$ is the same as in (a). }
\label{Fig:order-parameter}
\end{figure}

Let $R_1$ and $R_2$ be
the global order parameters of the layer I and II, respectively, defined by
$R_1e^{i\Psi_1}=\frac{1}{N}\sum_{j=1}^N e^{i\theta_{j,1}}$ and
$R_2e^{i\Psi_2}=\frac{1}{N}\sum_{j=1}^N e^{i\theta_{j,2}}$. In our simulations, $N=1,000$ and, for convenience, layer I
is fixed as a random ER network with average degree $\langle k_1\rangle=12$ \cite{BA:2002}, and we draw the set frequencies $\{\omega_{i,1}\}$
from a random homogeneous distribution in the range $[-1,1]$.
Instead, in the following, we will vary both the topology and the frequency distribution $g(\omega_{i,2})$ characterizing layer II.
First, we let it be an independent
random ER network with the same average
degree $\langle k_2\rangle=12$, and we let its frequencies $\{\omega_{i,2}\}$ be drawn from an
independent random homogeneous distribution in the range $[-1,1]$. Fig. \ref{Fig:order-parameter}(a) shows the dependence of
$R_1$ (``squares" and ``triangles") and $R_2$
(`circles" and ``stars") on $\lambda$ for the case of $f=1$, where the ``squares" and ``circles" (``triangles" and ``stars")
denote the forward (backward) transition. One clearly see that ES sets in both layers.
Once again, the inset of Fig. \ref{Fig:order-parameter}(a) shows the
dependence of $d$ on $f$, and, in analogy with Fig. \ref{Fig:single-network}, indicates the presence of a critical value $f_c$ for the setting
of an irreversible, hysteretic transition.

Second, we let the average degree $\langle k_2\rangle$ change from $12$ to $6$ while keeping all other
parameters unchanged. The results for  $f=1$ are now shown in Fig. \ref{Fig:order-parameter}(b), with the inset
reporting again the dependence of $d$ on $f$. Comparing Fig. \ref{Fig:order-parameter}(b) with (a), one see that once again
ES sets in, though the associated values of d and $f_c$ are, respectively, smaller and larger.

As a third step, we now let the frequency distribution $g(\omega_{i,2})$ change from a homogeneous to a
Lorentzian distribution $g(\omega)=\frac{1}{\pi}[\frac{\gamma}{(\omega-\omega_0)^2+\gamma^2}]$ with
central frequency $\omega_0=0$ and $\gamma$ (the half width at half maximum \cite{Matthews:1991}) equal to 0.5,
while keeping all other parameters as those of Fig. \ref{Fig:order-parameter}(a). A significant difference between
the homogeneous distribution and the Lorentzian distribution is that the former is homogeneous
for every $\omega$ while the latter is heterogeneous with an approximate power law on $\omega$.
Nonetheless, Fig. \ref{Fig:order-parameter}(c) (corresponding to $f=1$) and its inset clearly indicate the setting of ES.

Finally, we even change the topology of layer II from an ER network to a Barab\'{a}si-Albert (BA) network \cite{BA:2002}, while
keeping all other parameters as in the case of Fig. \ref{Fig:order-parameter}(a).
Notice that, in this latter situation, the topologies of the two layers are essentially different. Once again,
the results (reported in Fig. \ref{Fig:order-parameter}(d)) are similar to those of the three panels (a)-(c),
and demonstrate the existence of both
an hysteretic loop and a critical $f_c$ associated with the transition to synchronization.

\begin{figure}
\epsfig{figure=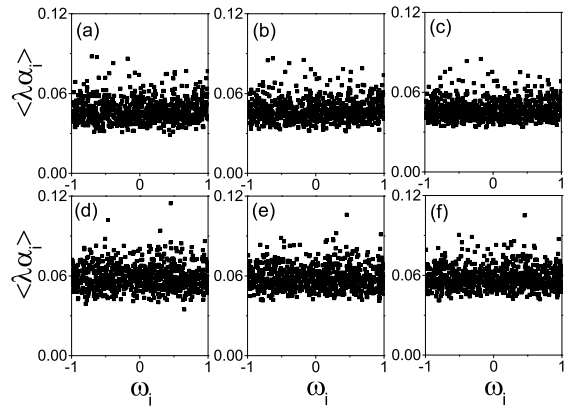,width=1.0\linewidth} \caption{Correlation properties just before the forward transition to synchronization.
Panels (a),(b) and (c) refer to the single network case. They corresponds to $f=0.8$, $\lambda=0.16$ (i.e. at the forward transition point to synchronization),
a single ER network with $1,000$ nodes, $<k>=12$, and a uniform frequency distribution in the range [-1; 1].
Panels (a),(b) and (c) report the average of $\lambda \alpha_i$ vs. $\omega_i$ over 5,000, 10,000 and 50,000 time steps, respectively.
Panels (d),(e) and (f) account, instead for the two-layer case, with $N=1,000$, $f=0.8$, $\lambda=0.2$
(once again at the forward transition point), and all other parameters as in the caption of Fig.
(\ref{Fig:order-parameter} (a)). These latter panels report actually the average of $\lambda \alpha_i$ vs. $\omega_i$ over 5,000, 10,000 and 50,000 time steps,
respectively. Exactly the same qualitative scenario (not shown) occur for the nodes in layer II).
}
\label{Fig:correlations}
\end{figure}

We now stop for a moment, and try to recall (at this stage) the remarkable conclusions that can be drawn from what already
reported so far. ES is a generic property of adaptive networks, as well as multilayer networks, as far as the coupling form
used in Eq. (\ref{Kuramoto1}) and Eq. (\ref{Kuramoto}) is taken into account.
This has two main implications: (i) it sharply contrasts to previous conclusions that a positive
correlation between the natural frequencies of oscillators and their effective couplings
is a necessary condition for ES; and (ii) the passage from a first- to a second-order transition is
here actually controlled by the coupled fraction $f$ of nodes for which adaptation is effective.
Let, therefore, us to speculate more on the correlation issue. The fraction $1-f$ of oscillators not affected by the adaptation mechanism is,
by definition and construction, not displaying {\it any form} of correlation between nodes' frequencies and coupling strength.
As for the faction $f$ of nodes which are, instead, affected by the adaptation rule, it is determinant to notice that the form used in Eq. (\ref{local-order1})
is actually {\it fully independent} on the actual instantaneous phase of the considered oscillators
(it, indeed, depends {\it only and solely} on the degree of
synchronization of the neighboring oscillators of each of these nodes). And, indeed, Fig. \ref{Fig:correlations}
show that {\it no correlation features at all} between the oscillators' frequencies and their coupling strengths are associated
(at the onset of ES in both single-layer and two-layers networks) neither in their short-,
nor in their intermediate-, nor even in their long-time averages (as compared with the time scale of their evolution).

In order to gather further information on the overall scenario, we move to some theoretical analysis, and
take the case of Fig. \ref{Fig:single-network}
with $f=1$ as an example. Substituting
Eq. (\ref{local-order1}) into Eq. (\ref{Kuramoto1}) one has
\begin{equation}
\dot{\theta}_{i}=\omega_{i}+\lambda r^2_ik_{i}\sin(\Psi-\theta_{i}) ,
\label{mean-field}
\end{equation}
where $\dot{\Psi}=\Omega$ is the group angular velocity. In the mean-field framework, $r_{i}=R$.
Letting $\Delta \theta_{i}=\theta_{i}-\Psi$, Eq. (\ref{mean-field}) becomes
\begin{equation}
\Delta\dot{\theta}_{i}=\omega_{i}-\Omega-\lambda R^2k_{i}\sin(\Delta\theta_{i}) .
\label{mean-field1}
\end{equation}
If $|\omega_{i}-\Omega|<\lambda R^2k_{i}$, then $\Delta\dot{\theta}_{i}$ reaches a fixed point defined by
$\sin(\Delta\theta_{i})=(\omega_{i}-\Omega)/\lambda R^2k_{i}$, indicating that the oscillator $i$ becomes
phase locked to the mean-field.
Otherwise, $\Delta\dot{\theta}_{i}$ never reaches a fixed point, indicating that
oscillator $i$ drifts at all time. Considering that the natural frequency distribution $g(\omega_{i})$
is here symmetric, we have that the average frequency $\Omega=0$. Thus, for the phase-locked
oscillators, one has
\begin{equation}
\Delta{\theta}_{i}=\arcsin(\frac{\omega_{i}}{\lambda R^2k_{i}}), \quad |\omega_{i}|\leq \lambda R^2k_{i} .
\label{mean-field2}
\end{equation}

Based on Eq. (\ref{mean-field2}), one can calculate the order parameter $R$. Noticing that
$R e^{i\Psi}=\frac{1}{N}\sum_{j=1}^N e^{i\theta_{j}}=\frac{1}{N\langle k\rangle}\sum_{j=1}^N k_je^{i\theta_{j}}$
and that the drifting oscillators do not contribute to $R$ \cite{Peron:2012,Skardal:2014}, one has
\begin{equation}
R=\frac{1}{N\langle k\rangle}\sum_{|\omega_{j}|\leq \lambda R^2k_{j}} k_{j}\cos(\Delta{\theta}_{j}) .
\label{mean-field3}
\end{equation}
Substituting Eq. (\ref{mean-field2}) into Eq. (\ref{mean-field3}) one obtains
\begin{equation}
R=\frac{1}{N\langle k\rangle}\sum_{|\omega_{j}|\leq \lambda R^2k_{j}} k_{j}\sqrt{1-(\frac{\omega_{j}}{\lambda R^2k_{j}})^2} .
\label{mean-field4}
\end{equation}
Replacing the summation over degrees by an integration, the contribution of the locked
oscillators to the order parameter in the thermodynamic limit is
\begin{equation}
R=\frac{1}{\langle k\rangle}\int_{|\omega|\leq \lambda R^2k} h(k,\omega)k\sqrt{1-(\frac{\omega}{\lambda R^2k})^2}d\omega dk ,
\label{mean-field5}
\end{equation}
where $h(k,\omega)$ is the joint distribution and can be written as $h(k,\omega)=P(k)g(\omega)$ with $P(k)$
being the degree distribution of the network.

If repeated for the case of Eqs. (\ref{Kuramoto}), the same treatment yields
\begin{eqnarray}
{\small R_1=\frac{1}{\langle k_1\rangle}\int_{C_{1}} h(k_{1},\omega_{1})k_{1}\sqrt{1-(\frac{\omega_{1}}{\lambda R_1R_2k_{1}})^2}d\omega_{1}dk_1},  \nonumber \\
{\small R_2=\frac{1}{\langle k_2\rangle}\int_{C_2} h(k_{2},\omega_{2})k_2\sqrt{1-(\frac{\omega_2}{\lambda R_1R_2k_2})^2}d\omega_2dk_2 },
\label{mean-field6}
\end{eqnarray}
where $C_{1,2} \equiv {|\omega_{1,2}|\leq \lambda R_1R_2k_{1,2}}$ are the integration domains,
$h(k_{1},\omega_{1})=P(k_{1})g(\omega_{1})$ and $h(k_{2},\omega_{2})=P(k_{2})g(\omega_{2})$, with
$P(k_{1})$ and $P(k_2)$ being the degree distributions of layer I and II, respectively.

\begin{figure}
\epsfig{figure=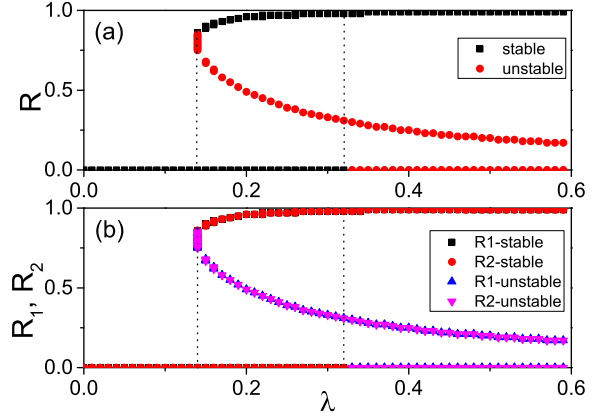,width=1.0\linewidth} \caption{(color
online). Analytical solutions for the order parameter $R$.
(a) $R$ vs. $\lambda$ for the single ER network of Eq. (\ref{Kuramoto1}) with $f=1$. $R$
is here calculated from Eq. (\ref{mean-field5}), and parameters are the same
as those in Fig. \ref{Fig:single-network}. (b) $R$ vs. $\lambda$ for the two-layered
ER network of Eqs. (\ref{Kuramoto}) with $f=1$. $R_1$ (``squares" and ``up triangles") and
$R_2$ (``circles" and ``down triangles") are evaluated from Eqs. (\ref{mean-field6}) and
the parameters are the same as in Fig. \ref{Fig:order-parameter}(a). In all cases, "dotted lines"
are just guides for the eye. }
\label{Fig:theory}
\end{figure}

Panels (a) and (b) of Fig. \ref{Fig:theory} report the solutions of Eqs. (\ref{mean-field5}) and
(\ref{mean-field6}), respectively. In both cases, it is easy to notice the presence of an unstable middle branch, which is
 responsible for the hysteretic loop associated to ES, and observed in Figs. \ref{Fig:single-network} and \ref{Fig:order-parameter}.

Our analytic results allow a better and deeper understanding of the intimate causes for ES, and
in particular of the microscopic mechanisms that are at the basis of the arousal of explosiveness
in the transition. Indeed, if one considers the usual Kuramoto model
$\dot{\theta}_{i}=\omega_{i}+\lambda\sum_{j=1}^NA_{ij}\sin(\theta_{j}-\theta_{i})$ for
the common second-order phase transition, and develops the same mean-field treatment, one obtains
that the formula for the order parameter is
\begin{equation}
R=\frac{1}{\langle k\rangle}\int_{|\omega|\leq \lambda Rk} h(k,\omega)k\sqrt{1-(\frac{\omega}{\lambda Rk})^2}d\omega dk.
\label{mean-general}
\end{equation}
Now, a distinctive difference between Eq. (\ref{mean-general}) and Eq. (\ref{mean-field5}) is that
the integration range $|\omega|\leq \lambda Rk$ in Eq. (\ref{mean-general}) is replaced by
$|\omega|\leq \lambda R^2k$ in Eq. (\ref{mean-field5}). Such a replacement results actually in the following,
remarkable, consequence. For the backward curves in Figs. \ref{Fig:single-network} and
\ref{Fig:order-parameter} one has $R\approx R^2\approx 1$, and thus the difference between
$R$ and $R^2$ is not so large. Instead, for the forward curves of the transition, one has $R\approx 0$ and thus the difference between $R$
and $R^2$ is there significant, i.e. $R^2$ will be much smaller than $R$. Notice, further, that the integration
domains in Eqs. (\ref{mean-general}) and (\ref{mean-field5}) actually determine the fraction of oscillators
belonging to the main synchronization cluster. In other words, the
larger synchronized clusters are forbidden to be formed in Eq. (\ref{mean-field5}), in analogy with
the suppressive rule recently discussed in Ref. \cite{Zhang:2014}. In details, in the usual case of a second-order transition,
the oscillators with closer natural frequencies will first form small
synchronized clusters, and then these clusters will gradually grow up and merge with the increase
of the coupling strength, up to eventually forming a giant cluster. On the contrary,
in the present case, the factor $R^2$ in the integration domain has the effect of actually {\it suppressing} the merging of
small synchronized clusters. Thus, with the increase of $\lambda$, more and more free oscillators
will be attracted to the distinct clusters, but these clusters are prevented from merging each other. Eventually, when
no more free oscillators are left, a discontinuous and abrupt behavior of $R$ will show up as the consequence of the sudden collapse of all
clusters, determining a first-order like transition.

Notice that the above discussions holds for the case $f=1$. When $f<1$, the oscillators can actually be divided
into two groups, the controlled fraction $f$ and the free fraction $1-f$. The oscillators in the
controlled fraction have a behavior similar to that of the case $f=1$, while those in the
fraction $1-f$ will behave similarly to Eq. (\ref{mean-general}). Considering that the
controlled group $f$ and the free group $1-f$ are in fact inter-connected, the behavior of the free part
$1-f$ will be influenced by the controlled part $f$, implying that the merging of small clusters
in the free part $1-f$ will be again suppressed. When $1>f>f_c$, this suppression mechanism will be the leading one,
and therefore the system will exhibit ES. On the contrary, when $f<f_c$, the suppression mechanism will not be strong
enough, and the merging behavior of small clusters in the free part $1-f$ will cause the occurrence of a
second-order smooth transition.

The most remarkable conclusion of our study is therefore that ES has, indeed, a microscopic root, but this root is essentially to be
found in a microscopic mechanism able to suppress the
formation of a giant synchronization cluster. While a positive correlation between the oscillators' natural frequencies and their
degrees \cite{Jesus:2011,Leyva:2012} or couplings strength \cite{Leyva:2013a,Zhang:2013,Leyva:2013b} has the effect of
suppressing at all the formation of any synchronization cluster, in the present case (i.e. in the absence of
any specific correlation features), the network nodes are initially able to form small {\it independent} synchronized clusters,
each one of them being able to further grow with the increase of the coupling strength,
and the suppression mechanism acts instead by impeding a merging process of the clusters.
We cannot therefore exclude that even other forms of realizing such a suppressive rule, originating from yet unrevealed
microscopic sources, would also determine the arousal of ES.

In conclusion, we reported on a novel framework for the setting of abrupt, explosive and irreversible transitions to synchronization
in adaptive and multilayer networks, where it is possible to observe ES even without the requirement of positive correlations between
natural frequencies and effective couplings of the networks' nodes. Our results are fully robust against large variations in the network
topologies and frequency distributions. Based on these findings and in contrast with the accepted state of knowledge on the subject,
we can safely conclude that the necessary condition for ES is, in fact, the existence of any microscopic suppressive rule
able to prevent (in a way or in another) the formation of a giant synchronization cluster.

Work partially supported by the NNSF of China under Grant No. 11135001 and 11375066,
Joriss project under Grant No. 78230050, 973 Program under Grant No. 2013CB834100, the Innovation
Program of Shanghai Municipal Education Commission grant No. 12ZZ043, and the Open Project Program
of State Key Laboratory of Theoretical Physics, Institute of Theoretical Physics, Chinese Academy of Sciences, China (No. Y4KF151CJ1).


\begin{references}


\bibitem{bocca}
S. Boccaletti, V. Latora, Y. Moreno, M. Chavez, and D.-U. Hwang, Phys. Rep. {\bf 424}, 175 (2006).

\bibitem{Dhamala:2013}
B. M. Adhikari, C. M. Epstein, and M. Dhamala,
Phys. Rev. E {\bf 88}, 030701(R) (2013).

\bibitem{Buldyrev:2010}
S. V. Buldyrev, R. Parshani, G. Paul, H. E. Stanley, and S. Havlin,
Nature {\bf 464}, 1025 (2010).

\bibitem{Huberman:1997}
B. A. Huberman and R. M. Lukose, Science {\bf 277}, 535 (1997).

\bibitem{pazo:2005}
D. Paz\'o, Phys. Rev. E {\bf 72}, 046211 (2005).

\bibitem{Jesus:2011}
J. G\'{o}mez-Garde\~{n}es, S. G\'{o}mez, A. Arenas and Y. Moreno,
Phys. Rev. Lett. {\bf 106}, 128701 (2011).

\bibitem{Leyva:2012}
I. Leyva, R. Sevilla-Escoboza, J. M. Buldu, I. Sendina-Nadal,
J. Gomez-Gardenes, A. Arenas, Y. Moreno, S. Gomez,
R. Jaimes-Reategui, and S. Boccaletti, Phys. Rev. Lett. {\bf 108}, 168702 (2012).

\bibitem{Peron:2012}
T. K. D. M. Peron and F. A. Rodrigues, Phys. Rev. E {\bf 86}, 056108 (2012);
ibid, {\bf 86}, 016102 (2012); B. C. Coutinho, A. V. Goltsev, S. N. Dorogovtsev, and J. F. F.
Mendes, Phys. Rev. E {\bf 87}, 032106 (2013); W. Liu, Y. Wu, J. Xiao, and M. Zhan, Europhys. Lett. {\bf 101}, 38002 (2013);
P. Ji, T. K. DM. Peron, P. J. Menck, F. A. Rodrigues, and J. Kurths,
Phys. Rev. Lett. {\bf 110}, 218701 (2013).

\bibitem{Leyva:2013a}
I. Leyva, I. Sendina-Nadal, J. A. Almendral, A. Navas, S. Olmi, and S. Boccaletti,
Phys. Rev. E {\bf 88}, 042808 (2013).

\bibitem{Zhang:2013}
X. Zhang, X. Hu, J. Kurths, and Z. Liu, Phys. Rev. E {\bf 88}, 010802 (R)
(2013).

\bibitem{Leyva:2013b}
I. Leyva, A. Navas, I. Sendina-Nadal, J. A. Almendral, J. M.
Buldu, M. Zanin, D. Papo, and S. Boccaletti, Sci. Rep. {\bf 3}, 1281 (2013).

\bibitem{Li:2013}
P. Li, K. Zhang, X. Xu, J. Zhang, and M. Small, Phys. Rev. E {\bf 87}, 042803 (2013);
L. Zhu, L. Tian, and D. Shi, Phys. Rev. E {\bf 88}, 042921 (2013).

\bibitem{Zhang:2014}
X. Zhang, Y. Zou, S. Boccaletti, and Z. Liu, Sci. Rep. {\bf 4}, 5200 (2014).

\bibitem{details} 
The stationary value of R is calculated by taking a long time average after the transient.

\bibitem{BA:2002}
R. Albert and A.-L. Barab\'{a}si, Rev. Mod. Phys.{\bf 74}, 47 (2002).

\bibitem{Parshani:2011}
R. Parshani, S. V. Buldyrev, and S. Havlin, Proc. Natl. Acad. Sci.{\bf 108}, 1007 (2011).

\bibitem{Matthews:1991}
P. C. Matthews, R. E. Mirollo and S. H. Strogatz, Physica D {\bf
52}, 293 (1991).

\bibitem{Skardal:2014}
P. S. Skardal and A. Arenas, Phys. Rev. E {\bf 89}, 062811 (2014).



\end{references}
\end{document}